\documentclass[prd, aps, showpacs]{revtex4}

\usepackage{latexsym, graphicx}

\begin{document}

\title{Disentanglement of two harmonic oscillators in relativistic motion}
\author{Shih-Yuin Lin}
\email{sylin@phys.cts.nthu.edu.tw}
\affiliation{Physics Division, National Center for Theoretical Science,
P.O. Box 2-131, Hsinchu 30013, Taiwan}
\author{Chung-Hsien Chou}
\email{chouch@mail.ncku.edu.tw}
\affiliation{Department of Physics, National Cheng Kung University}
\affiliation{Physics Division, National Center for Theoretical Sciences(South),
Tainan 701, Taiwan}
\author{B. L. Hu}
\email{blhu@umd.edu}
\affiliation{Joint Quantum Institute and
Maryland Center for Fundamental Physics, Department of Physics,
University of Maryland, College Park, Maryland 20742-4111, USA}
\affiliation{
Perimeter Institute for Theoretical Physics,
31 Caroline Street North, Waterloo, Ontario N2L 2Y5 Canada }
\date{28 March 2008 (v1); 19 December 2008 (v2); Phys. Rev. D {\bf 78}, 125025 (2008)}
\begin{abstract}
We study the dynamics of quantum entanglement between two Unruh-DeWitt
detectors, one stationary (Alice), and another uniformly accelerating
(Rob), with no direct interaction but coupled to a common quantum field
in (3+1)D Minkowski space. We find that for all cases studied the
initial entanglement between the detectors disappears in a finite time
(``sudden death"). After the moment of total disentanglement the
correlations between the two detectors remain nonzero until late times.
The relation between the disentanglement time and Rob's proper
acceleration is observer dependent. The larger the acceleration is,
the longer the disentanglement time in Alice's coordinate, but the
shorter in Rob's coordinate.                                                
\end{abstract}


\maketitle

\section{Introduction}

A causally disconnected (spacelike separated) pair of qubits or atoms
on the same time slice can be quantum correlated and entangled. 
This evokes the notion of ``nonlocality" as an innate feature of           
quantum entanglement, the precise meaning of which is a topic of             %
sustained interest and some controversy. When one examines the quantum 
entanglement across the event horizon of a black hole, the notion of 
``nonlocality" acquires an additional layer of meaning, pertaining not 
only to quantum correlations in ordinary (Minkowski) spacetime but           %
also to some nontrivial (global) spacetime structure.                      

When the atoms are coupled with quantum fields, the situation becomes
more interesting. Interaction with the quantum fields will induce
decoherence of the atoms and affect the entanglement between them. It
is known that the behavior of quantum entanglement is in general very
different from decoherence. For the dynamics of quantum entanglement
between two qubits different environmental settings could lead to
very different results. (Compare, e.g, \cite{YE04,FicTan06,ASH06}).
Here we add in another dimension of consideration, that arising from
nontrivial global structure of spacetime such as the existence of an
event horizon,  as in the spacetime of a black hole (Schwarzschild)
or a uniformly accelerated detector (UAD). Specifically, how quantum
entanglement between two detectors across the event horizon, one
stationary and another uniformly accelerating, would evolve in time and
how causality effects including that of retarded mutual influences
would play out in these processes. This is an important ingredient
for the establishment of relativistic quantum information theory.

Alsing and Milburn \cite{AM03} considered the quantum entanglement
between two detectors (a quantum object with internal degrees of
freedom), one is inertial (Alice) and the other is in relativistic
motion (Bob, but when in uniform acceleration, they call it Rob) .
Using the fidelity of the teleportation as a measure of entanglement,
they claimed that the entanglement is degraded in noninertial frames
due to the Unruh effect \cite{Unr76}.
In their treatment, both detectors are made of
cavities, and the qubits are constructed by using ``single particle
excitations of the Minkowski vacuum states in each of the cavities"
of a scalar field. However, as is pointed out by Sch\"utzhold and
Unruh \cite{SU05}, the introduction of cavities alters the boundary
conditions in the derivations of the Unruh effect \cite{Unr76, DeW79,
BD}, and the evolution of the field modes depends on the way the
cavity is accelerated. Furthermore, if Rob's cavity is stationary in
its local frame, then there is no particle creation inside, otherwise
one has to take into account the dynamical Casimir effect
\cite{SU05}.

Without the trouble associated with the cavity, Fuentes-Schuller and
Mann \cite{FSM05} consider free field modes in Minkowski space and
assume the inertial Alice and the accelerated Rob are observers
sensitive to different single modes. Suppose the quantum field is in
a maximally entangled state of these two modes and the uniformly
accelerated Rob is always in the right Rindler wedge, then from the
negativity of the observed reduced density matrix (which is obtained
by integrating out the mode in the left Rindler wedge originally
sensitive to Rob at rest but undetectable when Rob is accelerated),
Fuentes-Schuller and Mann claimed that the field state is {\it
always} entangled if Rob's acceleration is finite, and because of the
Unruh effect,  the larger Rob's acceleration, the smaller the
entanglement between the two modes. Note that the entanglement here
is not between qubits or anything localized in Alice or Rob's hands;
It is the entanglement between the two modes spread in the whole
Minkowski space and observed locally by Alice and Rob without
considering the back-reaction of the field on the detector. Note also
that both the fidelity in \cite{AM03} and the negativity in
\cite{FSM05} are independent of time. Thus there is no dynamics in
these two earlier results.

\subsection{Main Features in this Problem}
\label{feature}

In this paper we will consider a more realistic model with two
Unruh-DeWitt (UD) detectors (atoms), which are pointlike objects
with internal degrees of freedom described by harmonic
oscillators, moving in a quantum field: Alice is at rest and Rob
is in uniform acceleration. These two detectors are set to be
entangled initially, while the initial state of the field is the
Minkowski vacuum.

The model has the advantage that it encompasses the main issues of
interest yet simple enough because of its linearity to yield
analytical solutions over a full parameter range. For the case of a
single uniformly accelerated detector in a quantum field studied
previously \cite{LH2005, LH2006}, exact results are available in
closed form. Here we will apply similar techniques to study how
quantum entanglement evolves between two initially entangled UD
detectors interacting through the quantum field.

In our nonperturbative treatment the field will be evolving with the
detectors as a combined closed system. The back-reactions from
Alice's inertial detector and Rob's uniformly accelerated detector on
the quantum field are automatically included in a self-consistent
way. Moreover, our covariant formulation can account for fully
relativistic effects. The following effects or features are included
in our consideration:

1) {\it Unruh effect.} In calculating the two-point functions of the
detectors one can see that the uniformly accelerated detector (Rob)
would experience vacuum fluctuations different from those seen by the
inertial detector (Alice). It was shown in \cite{LH2006} that in the
Markovian regime the field state which looks like the vacuum for the
inertial detector behaves like a thermal state for the uniformly
accelerated one. This is the Unruh effect \cite{Unr76}.

2) {\it Causality and retardation of mutual influences.} The initial
information and the response from one detector to vacuum fluctuations
of the field will propagate outwards and reach the other detector in
finite time. But the influence on the other detector can in turn
propagate back to the original one, which creates memory effects.
These mutual influences should observe causality, and not propagate
faster than the speed of light. They are contained in the higher
order effects of quantum entanglement which can be calculated in our
formulation.

3) {\it Observer-dependence.}                                            
We learnt from the one-detector case that detectors with large
acceleration have noticeable changes in time only around $t=0$ because
significant time dilation will be seen by the field once $|t|$ is large
enough \cite{LH2007}. Therefore from the viewpoint of Alice and the
field most of the time the detector at Rob's place looks frozen.
On the other hand, in Rob's coordinate time Alice will take infinite
time to reach the event horizon, and Alice also looks frozen around
Rob's event horizon. These may alter the evolution of physical
quantities in their dependence on Rob's acceleration.

4) {\it Fiducial time.} While there is no simultaneity over space in
our relativistic model, there exists a specified time slice
(Minkowski time $t=0$ ) when we define the Hamiltonian and
an initial state (as a direct product of a squeezed state of two
detectors and the Minkowski vacuum of the field). At every moment of
time the physical reduced density matrix (RDM) for the two detectors
will be obtained by integrating out the degrees of freedom of the
field on that same time slice. One has to be cautious about whether       
the RDM depends on the time slicing or not.                               

\subsection{Key Issues of Interest}

The following key issues are addressed in this work:

1) {\it Disentanglement, ``sudden death" and entanglement revival.}
Yu and Eberly \cite{YE04} discovered that, unlike the decoherence
process, for two initially entangled qubits each placed in its own
reservoir completely detached from the other, the disentanglement
time can be finite, namely, quantum entanglement between these two
qubits can see a sudden death. Ficek and Tanas \cite{FicTan06} as
well as Anastopoulos, Shresta and Hu \cite{ASH06} studied the problem
where the two qubits interact with a common electromagnetic field.
The former authors while invoking the Born and Markov approximations
find the appearance of dark periods and revivals. The latter authors
treat the non-Markovian behavior without these approximations and
find a different behavior at short distances. In particular, for weak
coupling, they obtain analytic expressions for the dynamics of
entanglement at a range of spatial separation between the two qubits,
which cannot be obtained when the Born-Markov approximation is
imposed. We wish to investigate in the particular setup of our model
whether and how these distinct feature of ``death" and/or ``revival"
manifest in the dynamics of entanglement.

2) {\it Entanglement in different coordinates.}                              
Following the well-known recipe, measures of entanglement such as              %
logarithm negativity \cite{VW02} can be calculated in a new coordinate 
with a time slicing different from Minkowski times ({\it e.g.} Rindler 
time). We will study whether those measures of entanglement in a new 
coordinate can be interpreted as the degree of entanglement in Rob's 
clock (Rindler time). If yes, what is the difference between the               %
entanglement dynamics in different coordinates?                              

3) {\it Spatial separation between two detectors.}
How does the entanglement vary with the spatial separation $d$ between        
the two qubits? In (3+1)D (dimension) the mutual influences                   
on mode functions are proportional to $d^{-1}$ so it is quite small
for large $d$ even if the coupling is not ultraweak. Still, it                
is of interest to see whether the mutual influences suppress or 
enhance quantum entanglement,                                                   
as compared to those from local vacuum fluctuations at each detector.         

These are some interesting new issues which will be expounded in our
present study.

\subsection{Summary of Our Findings}

The results from our calculations show that the interaction between
entangled UD detectors and the field does induce quantum
disentanglement between the two detectors. We found that the
disentanglement time is finite in all cases studied, namely, there is
no residual entanglement at late times for two spatially separated
detectors, one stationary and another uniformly accelerating, in
(3+1)D Minkowski space. Around the moment of full disentanglement
there may be some short-time revival of entanglement within a                 
few periods of oscillations of the detectors (equal                           
to the inverse of their natural frequency $\Omega$). But there is no
entanglement generated at times much longer than $O(1/\Omega)$.

In the ultraweak-coupling limit, the leading-order behavior of
quantum entanglement in Minkowski time is independent of Rob's proper
acceleration $a$. When $a$ gets sufficiently large, the
disentanglement time from Alice's view would be longer for a larger
$a$. From Rob's view, however, the larger $a$ is the shorter the
disentanglement time.                                                         
Finally, in the strong-coupling regime, the strong impact of vacuum           
fluctuations experienced locally by each detector destroys their
entanglement right after the coupling is switched on.

\subsection{Outline of this Paper}                                        

This paper is organized as follows. In Sec. II we describe the setup 
of the problem, introduce the model and  describe the measure of       
quantum entanglement we use. In Sec. III we present our
calculations in the Heisenberg picture of the evolution of the
operators and the two-point functions of the detectors. In Sec. IV
we illustrate the results in the ultraweak-coupling limit and beyond.
In Sec. V we present a discussion on a few key issues: (a)                   
infinite disentanglement time in Markovian limit, (b) entanglement             %
and correlation, (c) coordinate dependence, (d) detector-detector            
entanglement vs detector-field entanglement, e) the relation between
the degree of quantum entanglement and the spatial separation of two
detectors, and f) how generic the features illustrated by our
results are. In Appendix A, we show the analytic form of the mode
functions, while in Appendix B, the result of the case with two
inertial detectors weakly coupled with a thermal bath is given for
comparison.

\section{The model}

\begin{figure}
\includegraphics[width=8cm]{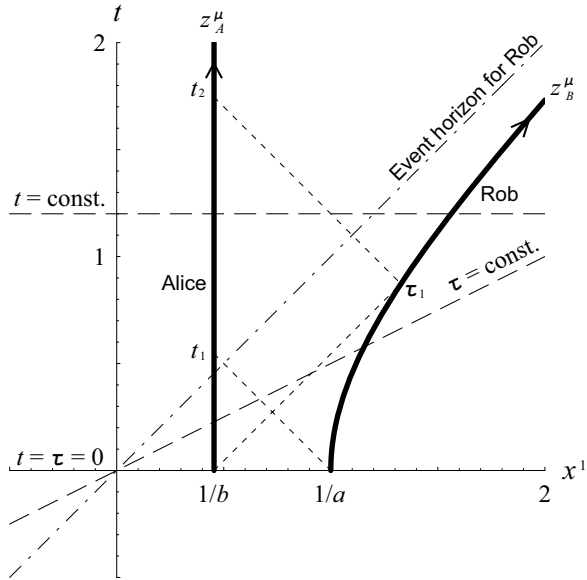}
\caption{Alice is at rest along the world line $z_A^\mu(t)=(t,1/b,0,0)$
and Rob is uniformly accelerated along the trajectory
$z_B^\mu(\tau)=(a^{-1}\sinh a\tau, a^{-1}\cosh a\tau,0,0)$, so the
null hypersurface $t = x^1$ is Rob's event horizon. The initial
state of the combined system is defined on the hypersurface $t=\tau=0$,
when the coupling between the detectors and the field is turned on. In
this plot we choose $b=2.2$ and $a=1$ so $b>2a$.} \label{setup}
\end{figure}

Consider two Unruh-DeWitt detectors moving in (3+1)-dimensional Minkowski     
space. The total action is given by\cite{LH2005}
\begin{eqnarray}
  S &=& -\int d^4 x \sqrt{-g} {1\over 2}\partial_\mu\Phi                      
    \partial^\mu\Phi \nonumber\\
    & & + \int d\tau_A {1\over 2}\left[\left(\partial_{\tau_A}Q_A\right)^2
    -\Omega_{0}^2 Q_A^2\right] +  \int d\tau_B{1\over 2}\left[\left(
    \partial_{\tau_B} Q_B\right)^2-\Omega_{0}^2 Q_B^2\right]\nonumber\\
    & & + \lambda_0\int d^4 x \Phi (x)\left[ \int d\tau_A Q_A(\tau_A)
  \delta^4\left(x^{\mu}-z_A^{\mu}(\tau_A)\right) + \int d\tau_B Q_B(\tau_B)
  \delta^4\left(x^{\mu}-z_B^{\mu}(\tau_B)\right)\right], \label{Stot1}
\end{eqnarray}
where $g_{\mu\nu} = {\rm diag}(-1,1,1,1)$,                                    
$Q_A$ and $Q_B$ are the internal degrees of freedom of Alice
and Rob, assumed to be two identical harmonic oscillators with mass
$m_0 =1$, bare natural frequency $\Omega_0$, and the same local
time resolution (so their cutoffs $\Lambda_0$ and $\Lambda_1$ in the
two-point functions \cite{LH2005} are the same). $\tau_A$ and
$\tau_B$ are proper times for $Q_A$ and $Q_B$, respectively. The
scalar field $\Phi$ is assumed to be massless, and $\lambda_0$ is the
coupling constant. Alice is at rest along the world line
$z_A^\mu(t)=(t,1/b,0,0)$ ($\tau_A = t$) and Rob is uniformly accelerated
along the trajectory $z_B^\mu(\tau)=( a^{-1}\sinh a\tau, a^{-1}\cosh a\tau,
0,0)$ ($\tau_B=\tau$) with proper acceleration $a$.
For simplicity we consider the cases with $b > 2a$ (see Fig. \ref{setup}).

Note that our model ($\ref{Stot1}$) is different from those
for quantum Brownian motion (QBM) of two harmonic oscillators (2HO) in
\cite{CYH07, PR07} (and in \cite{AZ07} in the rotating-wave approximation),
where the cases considered are analogous to two Unruh-DeWitt detectors at
the same spatial point. This is why there is no retarded mutual influence
between 2HO's in \cite{CYH07, PR07, AZ07}.
Also here the spectrum of quantum field fluctuations felt by Alice
and by Rob are different, while the vacuum fluctuations look the same
for the 2HO's in \cite{CYH07, PR07, AZ07}.

Suppose the coupling between the detectors and the field is turned on
at $t=\tau=0$, when the initial state of the combined system is a
direct product of a quantum state $\left|\right. q_A, q_B
\left.\right>$ for Alice and Rob's detectors $Q_A$ and $Q_B$ and the
Minkowski vacuum $\left|\right. 0_M \left.\right>$ for the field
$\Phi$, namely,
\begin{equation}
  \left|\right. \psi(0)\left.\right> =
  \left|\right. q^{}_A, q^{}_B\left.\right> \otimes
  \left|\right. 0_M \left.\right>. \label{initstat}
\end{equation}
Here $\left|\right. q^{}_A, q^{}_B \left.\right>$ is taken to be a squeezed
Gaussian state with minimal uncertainty, represented in the Wigner function as
\begin{equation}
  \rho(Q_A,P_A,Q_B,P_B) = {1\over \pi^2\hbar^2}\exp -{1\over 2}\left[              
  {\beta^2\over\hbar^2}\left( Q_A + Q_B\right)^2 +
  {1\over \alpha^2}\left( Q_A - Q_B\right)^2 +
  {\alpha^2\over\hbar^2}\left( P_A - P_B\right)^2 +
  {1\over \beta^2}\left( P_A + P_B\right)^2 \right],
\label{initGauss}
\end{equation}
in which $Q_A$ and $Q_B$ can be entangled by properly choosing the
parameters $\alpha$ and $\beta$.

Define the two-point correlation matrix $V$ with elements
\begin{equation}
  V_{\mu\nu}(t,\tau)=\left<\right.{\cal R}_\mu ,{\cal R}_\nu \left.\right>
    \equiv {1\over 2}\left<\right. \left( {\cal R}_\mu {\cal R}_\nu +
    {\cal R}_\nu {\cal R}_\mu \right) \left.\right> \label{CorrMtx}
\end{equation}
where ${\cal R}_\mu = (Q_B(\tau), P_B(\tau), Q_A(t), P_A(t))$,
$\mu, \nu = 1,2,3,4$.
The partial transpose of ${\bf V}$ is ${\bf V}^{PT} = {\bf \Lambda V
\Lambda}$, where ${\bf \Lambda} ={\rm diag}( 1,1,1,-1)$. Starting with
the Gaussian initial state $(\ref{initGauss})$, the reduced density
matrix or Wigner function of the two detectors is always Gaussian by virtue
of the linearity of our model. Therefore Alice's detector $Q_A$ and Rob's
detector $Q_B$ is entangled on time slice $t$ if and only if \cite{Si00}
\begin{equation}
  \Sigma(t, \tau=a^{-1}\sinh^{-1}at) \equiv \det \left[ {\bf V}^{PT}+
  i{\hbar\over 2}{\bf M} \right] < 0  \label{Entcond}
\end{equation}
where
\begin{equation}
  {\bf M} \equiv \left( \begin{array}{cccc}
    0 & 1 & 0 & 0 \\
    -1 & 0 & 0 & 0 \\
    0 & 0 & 0 & 1 \\
    0 & 0 & -1 & 0 \end{array} \right)
\end{equation}
is a symplectic matrix. We find that when $\Sigma \le 0$ the behavior of
$\Sigma$ is quite similar to the behavior of the negative eigenvalues
which are connected to the logarithm negativity \cite{VW02}. Thus the
value of $\Sigma$ itself is a good indicator of the degree of
entanglement, at least in this specific model. Since it is relatively
easy to obtain the analytic form of $\Sigma(t)$ in the weak-coupling
limit, we will calculate $\Sigma$ rather than the logarithm negativity
to determine the disentanglement time analytically.

Below we calculate the two-point functions for matrix $V$. The uncertainty
relation $ \det [ {\bf V} + i\hbar{\bf M}/2  ] \ge 0 $
can serve as a double-check \cite{Si00}.

\section{Evolution of operators and correlators}

\subsection{Evolution of Operators}

In the Heisenberg picture \cite{LH2005, LH2006}, the operators evolve
as
\begin{eqnarray}
  \hat{Q}_i(\tau_i) &=& \sqrt{\hbar\over 2\Omega_r}\sum_j\left[
    q_i^{(j)}(\tau_i)\hat{a}_j^{}+q_i^{(j)*}(\tau_i)\hat{a}_j^\dagger \right]
   +\int {d^3 k\over (2\pi)^3}\sqrt{\hbar\over 2\omega}
    \left[q_i^{(+)}(\tau_i,{\bf k})\hat{b}_{\bf k} +
    q_i^{(-)}(\tau_i,{\bf k})\hat{b}_{\bf k}^\dagger\right], \\
  \hat{\Phi}(x) &=& \sqrt{\hbar\over 2\Omega_r}\sum_j\left[
    f^{(j)}(x)\hat{a}_j^{}+f^{(j)*}(x)\hat{a}_j^\dagger \right]
    +\int {d^3 k\over (2\pi)^3}
    \sqrt{\hbar\over 2\omega}\left[f^{(+)}(x,{\bf k})\hat{b}_{\bf k}
    +f^{(-)}(x,{\bf k})\hat{b}_{\bf k}^\dagger\right],
\end{eqnarray}
with $i,j = A,B$, $\tau_A = t$, $\tau_B=\tau$. $q_i^{(j)}$, $q_i^{(\pm)}$,
$f^{(j)}$ and $f^{(\pm)}$ are the (c-number) mode functions.
The conjugate momenta are $\hat{P}_A(t) =\partial_t\hat{Q}_A(t)$,
$\hat{P}_B(\tau) =\partial_\tau \hat{Q}_B (\tau)$, and $\hat{\Pi}(x) =
\partial_t\hat{\Phi}(x)$. The Heisenberg equations of motion for
operators imply
\begin{eqnarray}
    \left( \partial_{\tau_i}^2 + \Omega_0^2\right)q_i^{(j)}(\tau_i) &=&
      \lambda_0 f^{(j)}(z_i^\mu (\tau_i)), \label{eomqAB1}\\
    \left( \partial_t^2 - \nabla^2 \right)f^{(j)}(x) &=& \lambda_0
      \left[\int_0^{\infty} dt\, q_A^{(j)}\delta^4(x -z^{}_A(t))
     +\int_0^{\infty} d\tau\, q_B^{(j)}\delta^4 (x-z^{}_B(\tau)) \right],
     \label{feAB1}\\
  \left(\partial_{\tau_i}^2 + \Omega_0^2\right)q_i^{(+)}(\tau_i,{\bf k}) &=&
      \lambda_0 f^{(+)}(z_i^\mu(\tau_i), {\bf k}), \label{eomq+1} \\
    \left( \partial_t^2 - \nabla^2 \right)f^{(+)}(x,{\bf k}) &=& \lambda_0
      \left[\int_0^{\infty} dt\, q_A^{(+)}(t,{\bf k})\delta^4(x -z^{}_A(t))
     +\int_0^{\infty}d\tau\,q_B^{(+)}(\tau,{\bf k})\delta^4(x-z^{}_B(\tau))
     \right] \label{fe+1} ,
\end{eqnarray}
which have the same appearance as the corresponding classical
dynamical equations. $f^{(j)}$ and $f^{(+)}$ look like classical
fields generated by two pointlike sources at $z^{}_A$ and $z^{}_B$.
Solving the field equations $(\ref{feAB1})$ and $(\ref{fe+1})$, one
obtains $f^{(j)}$ and $f^{(+)}$ related to $q_i^{(j)}$ and
$q_i^{(+)}$ by the retarded Green's functions of the field. Inserting
them into the equations of motion $(\ref{eomqAB1})$ and
$(\ref{eomq+1})$  one obtains the solutions of $q_i^{(j)}$ and
$q_i^{(+)}$. However, the self-field induced by $q_i^{(j)}$ and
$q_i^{(+)}$ diverge at the positions of the two detectors, so one has to
introduce cutoffs to handle them. Assuming Alice and Rob have the
same frequency cutoffs in their local frame, one can do the same
renormalization on frequency and obtain the effective equations of
motion under the influence of field \cite{LH2005}:
\begin{eqnarray}
  \left( \partial_t^2 +2\gamma\partial_t + \Omega_r^2 \right)q_A^{(j)}(t)
    &=& {\lambda_0^2\over 2\pi }{\theta\left[\tau_-(z_A(t))\right]\over
    a X(z_A(t))} q_B^{(j)}\left( \tau_-(z_A^\mu (t)) \right),
    \label{eomqA2} \\
  \left(\partial_\tau^2 +2\gamma\partial_\tau +\Omega_r^2\right)
    q_B^{(j)}(\tau) &=& {\lambda_0^2\over 4\pi}{\theta\left[ z_B^0(\tau)
    -R(z_B(\tau))\right]\over R(z_B(\tau))} q_A^{(j)}\left( z_B^0(\tau)
    -R(z_B(\tau)) \right), \label{eomqB2}\\
  \left( \partial_t^2 +2\gamma\partial_t + \Omega_r^2 \right)
    q_A^{(+)}(t,{\bf k}) &=& \lambda_0 f_0^{(+)}(z_A(t),{\bf k})+
    {\lambda_0^2\over 2\pi}{\theta\left[\tau_-(z_A(t))\right]\over
    a X(z_A(t))} q_B^{(+)}\left( \tau_-(z_A^\mu (t)), {\bf k} \right),
    \label{eomqA+2} \\
  \left(\partial_\tau^2 +2\gamma\partial_\tau +\Omega_r^2\right)
    q_B^{(+)}(\tau, {\bf k}) &=& \lambda_0 f_0^{(+)}(z_B(\tau),{\bf k})+
    {\lambda_0^2\over 4\pi} {\theta\left[ z_B^0(\tau)-R(z_B(\tau))\right]
    \over R(z_B(\tau))} q_A^{(+)}\left( z_B^0(\tau)-R(z_B(\tau)), {\bf k}
    \right), \label{eomqB+2}
\end{eqnarray}
where $\Omega_r \equiv \sqrt{\Omega^{2} + \gamma^{2}}$ is the
renormalized frequency, $\gamma \equiv \lambda_0^2 /8\pi$, and
\begin{eqnarray}
  f_0^{(+)}(x,{\bf k}) &\equiv& e^{-i \omega t + i{\bf k\cdot x}},\\
  R(x) &\equiv& \sqrt{\left(x_1-b^{-1}\right)^2 + \rho^2 }, \\
  X(x) &\equiv& \sqrt{\left( - UV+\rho^2+a^{-2}\right)^2 +4a^{-2}UV},\\
  \tau_-(x) &\equiv& -{1\over a} \ln {a\over 2|V|}
    \left( X-UV +\rho^2+ {1\over a^2}\right),
\end{eqnarray}
with $U=t-x_1$, $V=t+x_1$, $\omega=|{\bf k}|$, and
$\rho^2=x_2^2+x_3^2$. Here one can see that $q_A$ and $q_B$ are
causally linked.

The solutions of $q_A$ and $q_B$'s satisfying the initial conditions
\begin{eqnarray}
 && f^{(+)}(0,{\bf x};{\bf k}) = 
   e^{i{\bf k\cdot x}},\;\;\; \partial_t f^{(+)}(0,{\bf x};{\bf k})=
   -i\omega e^{i{\bf k\cdot x}}, \label{IC+}\\
 && q_A^{(A)}(0) = q_B^{(B)}(0)=1, \;\;\; \partial_t{q}_A^{(A)}(0)=
 \partial_\tau {q}_B^{(B)}(0)= -i\Omega_r, \label{ICa}
\end{eqnarray}
and $f^{(j)} (0,{\bf x}) =\partial_t f^{(j)} (0,{\bf x}) =
q_j^{(+)}(0;{\bf k})= \partial_{\tau_j} q_j^{(+)}(0;{\bf k}) =
q_A^{(B)}(0) = \partial_t{q}_A^{(B)}(0) = q_B^{(A)}(0)=
\partial_\tau {q}_B^{(A)}(0)=0$, are listed in Appendix A.

\subsection{Two-Point functions of detectors}

When sandwiched by the initial state $(\ref{initstat})$, the two-point
functions split into
\begin{equation}
\left<\right.{\cal R}_\mu ,{\cal R}_\nu \left.\right> =
 \left<\right.{\cal R}_\mu ,{\cal R}_\nu \left.\right>_{\rm v} +
 \left<\right.{\cal R}_\mu ,{\cal R}_\nu \left.\right>_{\rm a}
\end{equation}
where
\begin{eqnarray}
  \left<\right.{\cal R}_\mu ,{\cal R}_\nu \left.\right>_{\rm v} &=&
    {1\over 2}\left<\right. 0_M | \left( {\cal R}_\mu {\cal R}_\nu +
    {\cal R}_\nu {\cal R}_\mu \right) | 0_M \left.\right> \nonumber\\
  &=& {\rm Re} \int {\hbar d^3 k\over (2\pi)^3 2\omega}
    r_\mu^{(+)}(t_\mu , {\bf k})r_\nu^{(-)}(t_\nu ,{\bf k}),\label{RRv}\\
  \left<\right.{\cal R}_\mu ,{\cal R}_\nu \left.\right>_{\rm a} &=&
    {1\over 2}\left<\right.q^{}_A, q^{}_B | \left( {\cal R}_\mu {\cal R}_\nu +
    {\cal R}_\nu {\cal R}_\mu \right) | q^{}_A, q^{}_B\left.\right>
  \nonumber\\ &=& {1\over 4}
    \left\{  \hbar^2 \beta^{-2}{\rm Re}\left(r_\mu^{(A)}
    +r_\mu^{(B)}\right){\rm Re}\left(r_\nu^{(A)}+r_\nu^{(B)}\right) +
    \alpha^2 {\rm Re}\left(r_\mu^{(A)}-r_\mu^{(B)}\right){\rm Re}
    \left(r_\nu^{(A)}-r_\nu^{(B)}\right)+\right.\nonumber\\
    & &\left.\Omega_r^{-2}\left[ \beta^2 {\rm Im}\left(r_\mu^{(A)}
    +r_\mu^{(B)}\right){\rm Im}\left(r_\nu^{(A)}+r_\nu^{(B)}\right)
    + \hbar^2\alpha^{-2}{\rm Im}\left(r_\mu^{(A)}-r_\mu^{(B)}\right)
    {\rm Im}\left(r_\nu^{(A)}-r_\nu^{(B)}\right)\right]\right\} \label{RRa},
\end{eqnarray}
where $r_\mu^{(j)} = (q_B^{(j)}, p_B^{(j)}, q_A^{(j)}, p_A^{(j)})$ and
$p_i^{(j)} = \partial_{\tau_i} q_i^{(j)}$.
Substituting Eqs. $(\ref{qAA})$-$(\ref{qBp})$, one obtains the above
two-point functions straightforwardly. The only complication is that the
integration over $k$ space in ($\ref{RRv}$) may diverge, so one has to
introduce additional frequency cutoffs corresponding to the time-resolution
of the detector ($\Lambda_1$) and the time scale of switching on the
interaction ($\Lambda_0$) \cite{LH2005, LH2006}. After doing this, one has,
for example,
\begin{equation}
  \left<\right.Q_B^2(\tau) \left.\right>_{\rm v} =
  \left<\right.Q_B^2(\tau) \left.\right>_{\rm v}^{(0)} +
  \theta\left(\tau -\tau_1\right) \left<\right.Q_B^2(\tau)
  \left.\right>_{\rm v}^{(1)},
\end{equation}
where $\tau_1 \equiv a^{-1}\ln(b/a)$ (see Fig. \ref{setup}),
$\left<\right. Q_B^2 \left. \right>_{\rm v}^{(0)}$ is the two-point
function of a single uniformly accelerated detector (expressions of
$\left<\right.Q_B^2\left.\right>_{\rm v}^{(0)}$,
$\left<\right.P_B^2\left.\right>_{\rm v}^{(0)}$,
$\left<\right.Q_A^2\left.\right>_{\rm v}^{(0)}$ and
$\left<\right.Q_B^2\left.\right>_{\rm v}^{(0)}$ have been listed
in Appendix A of Ref. \cite{LH2006}), while the
higher order correction reads
\begin{eqnarray}
  \left<\right.Q_B^2(\tau) \left.\right>_{\rm v}^{(1)} &=&
  {4\gamma\over\Omega} \int_{\tau_1}^\tau d\tau'
    {e^{-r(\tau-\tau')}\sin\Omega(\tau-\tau')\over
    {1\over a}\cosh a\tau' -{1\over b} }
    \left<\right.Q_B(\tau), Q_A ( b^{-1}-a^{-1}e^{-a\tau'})
    \left.\right>_{\rm v}^{(0)}\nonumber\\
  &+& {4\gamma^2\over\Omega^2} \int_{\tau_1}^\tau d\tau'
    {e^{-r(\tau-\tau')}\sin\Omega(\tau-\tau')\over {1\over a}
    \cosh a\tau' -{1\over b} } \int_{\tau_1}^\tau d\tau''
    {e^{-r(\tau-\tau'')}\sin\Omega(\tau-\tau'')\over
    {1\over a}\cosh a\tau'' -{1\over b} } \times \nonumber\\
  & & \,\,\,\,\,\,\,\,\,\,\left<\right.Q_A ( b^{-1}-a^{-1}e^{-a\tau'}),
    Q_A ( b^{-1}-a^{-1}e^{-a\tau''}) \left.\right>_{\rm v}^{(0)}.
\end{eqnarray}
Calculating the cross correlations $\left<\right.{\cal R}_A(t),{\cal
R}_B(\tau) \left. \right>_{\rm v}^{(0)}$ (${\cal R}=P,Q$) is also
straightforward, though one has to be careful about the contours of
integration of each term on the complex plane of $\kappa$ (see
\cite{LH2005}). Note that $\Lambda_1$ is present only in
$\left<\right. P_A^2 \left. \right>_{\rm v}$ and $\left<\right. P_B^2
\left.\right>_{\rm v}$.

\section{Disentanglement dynamics}

\subsection{ultraweak-coupling limit}

In the ultraweak-coupling limit ($\gamma\Lambda_1 \ll a, \Omega$),
the corrections due to the retarded  mutual influences in mode
functions $(\ref{qAA})$-$(\ref{qBp})$ are $O(\gamma)$ and suppressed,
while the cross correlations $\left<\right. {\cal R}_A, {\cal R}_B
\left.\right>_{\rm v}$ (${\cal R}=P,Q$) accounting for the response
to the vacuum fluctuations are negligible.
The two-point functions $\left<\right. .. \left.\right>_{\rm a}$ behave
like $(\ref{Qj2aweak})$-$(\ref{QAPBaweak})$, and
\begin{eqnarray}
   \left<\right. Q_A^2(t)\left.\right>_{\rm v} &\approx&
     {\hbar\over 2\Omega} \left( 1-e^{-2\gamma t}\right),\\
   \left<\right. Q_B^2(\tau )\left.\right>_{\rm v} &\approx& {\hbar\over
     2\Omega}\coth{\pi\Omega\over a}\left( 1-e^{-2\gamma\tau}\right),
\end{eqnarray}
and $\left<\right. P_j^2(\tau_j)\left.\right>_{\rm v} \approx
\Omega^2 \left<\right. Q_j^2(\tau_j)\left.\right>_{\rm v}$, $j=A,B$.
It is straightforward to calculate $\Sigma$ and determine the
separability of Alice and Rob by inserting these two-point functions
into ($\ref{Entcond}$).

\subsubsection{Evolution of entanglement in Alice's proper time
(Minkowski time)} \label{weakMin}

\begin{figure}
\includegraphics[width=18cm]{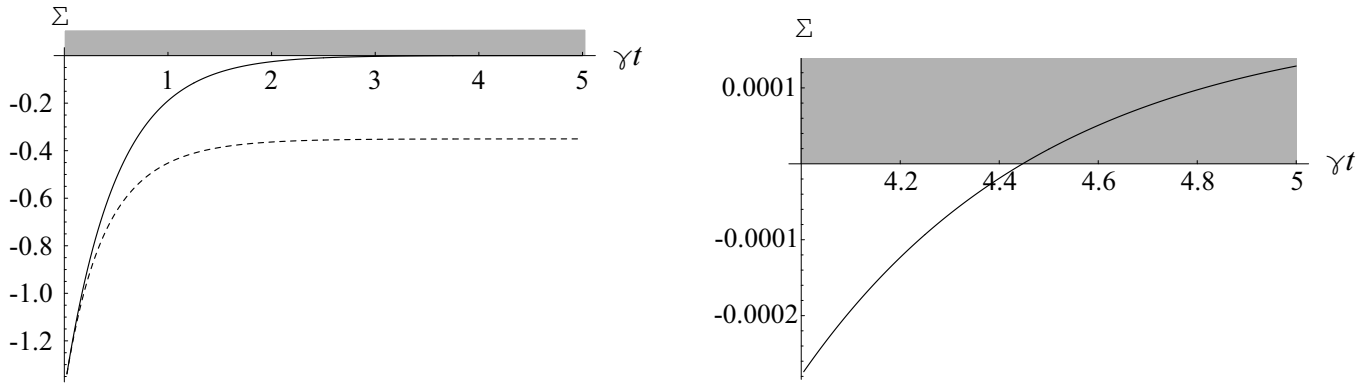}
\caption{(Left) The behavior of $\Sigma$ (solid curve) defined in
$(\ref{Entcond})$ can be approximated by $(\ref{Siga})$ in the
ultraweak-coupling limit with $a$ sufficiently large. Two detectors
are separable when $\Sigma \ge 0$ (shaded zone), otherwise entangled.
Here $\gamma=10^{-5}$, $a=0.1$, $b=0.201$, $\alpha= 1.1$, $\beta=4.5$,
$\Lambda_0=\Lambda_1=100$, $\Omega=2.3$ and $\hbar=1$. By comparison
the dotted curve is $\Sigma$ with all $\left<\right. ..\left.
\right>_{\rm v}$ set to be zero. Both curves will become
positive at late times. (Right) Looking more closely, one finds
that two detectors are separable, or totally disentangled, after
$\gamma t \approx 4.44$.} 
\label{AREt}
\end{figure}

At the initial moment $t=0$, the initial Gaussian state
$(\ref{initGauss})$ has
\begin{equation}
  \Sigma \approx -{\hbar^2 \over 16\alpha^2\beta^2} \left(
    \hbar^2 -\alpha^2\beta^2\right)^2 .
\end{equation}
Thus for all $\alpha^2\beta^2\not= \hbar^2$, two detectors are entangled.
Below we will see that quantum entanglement will, however, vanish at a
finite disentanglement time (``sudden death"), which is usually different
from the decoherence time scale $\gamma^{-1}$.

For proper acceleration $a$ sufficiently large such that                  
$\tau = a^{-1}\sinh^{-1}at \ll t$ for $t$ large enough, one has 
$e^{\gamma(t+\tau)}\approx e^{\gamma t}$ and
\begin{equation}
  \Sigma \approx -{\hbar^2\over 16\alpha^2\beta^2}\left( \hbar^2
  -\alpha^2 \beta^2\right)^2 e^{-2\gamma t}.
\label{Siga}
\end{equation}
An example is illustrated in Fig. \ref{AREt}. The disentanglement
time of this case looks infinite. However, the correction of the next
order will shift the curve of $\Sigma$ upward and make the
disentanglement time finite. When $\Lambda_1$ is large, such a
correction is dominated by $\Lambda_1$ terms in $\left<\right. P_j^2
\left.\right>_{\rm v}$:
\begin{eqnarray}
  \left<\right. P_A^2 \left.\right>_{\rm v} &\approx& {\hbar\over 2}\Omega
    \left( 1- e^{-2\gamma t}\right) +{2\over \pi}\hbar \gamma \Lambda_1,\\
  \left<\right. P_B^2 \left.\right>_{\rm v} &\approx& {\hbar\over 2}\Omega
    \coth{\pi\Omega\over a}\left( 1- e^{-2\gamma \tau}\right)
    +{2\over \pi}\hbar \gamma \Lambda_1
\end{eqnarray}
[see Eqs.(A4) and (A10) in Ref.\cite{LH2006}]. For $t_{dE}\gg 1$, this
correction yields
\begin{equation}
  \Sigma \approx -{\hbar^2\over 16\alpha^2\beta^2}\left( \hbar^2
  -\alpha^2 \beta^2\right)^2 e^{-2\gamma t} +
  {\hbar^2 \gamma\Lambda_1 \over 16\pi \alpha^2\beta^2}\left( \hbar^2
  -\alpha^2 \beta^2\right)^2,
\label{Siga1}
\end{equation}
around $t\approx t_{dE}$, which gives
\begin{equation}
  t_{dE} \approx {1\over 2\gamma}\ln {\pi\Omega\over \gamma\Lambda_1}.
\end{equation}
In Fig.\ref{AREt}, one has $\gamma t_{dE}\approx 4.44$.

Note that $\Sigma$ is insensitive to $a$ here. This implies that, to
leading order in the ultraweak-coupling approximation, Unruh effect
does little to the disentanglement between Alice and Rob {\it from
the view of Alice}.

\subsubsection{Two inertial detectors in Minkowski vacuum}
\label{2inertMin}

\begin{figure}
\includegraphics[width=18cm]{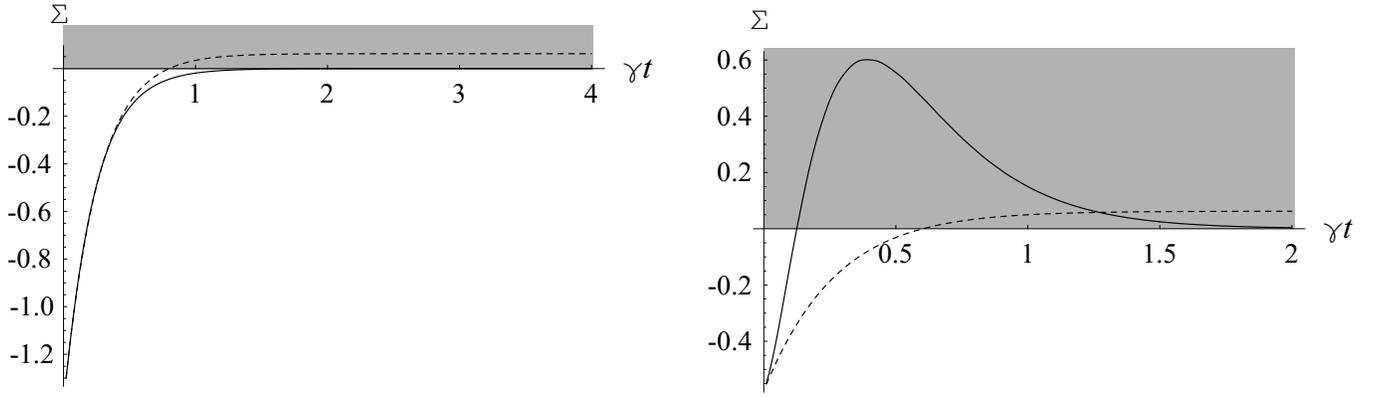}
\caption{ When $a\to 0$, the behavior of $\Sigma$ in the weak-coupling
limit can be described by $(\ref{Sig0})$.
Here $(\alpha,\beta) = (1.1, 4.5)$ for the left plot ($Z_4 < 0$) and
$(1.5, 0.2)$ for the right plot ($Z_4 > 0$). The solid curves represent
$\Sigma$ in its totality, while the dotted curves represent $\Sigma$
with all $\left<\right. .. \left.\right>_{\rm v}$ set to be zero.}
\label{AREta0}
\end{figure}

For $a\to 0$, $\coth (\pi\Omega/a) \to 1$, $\tau\to t$, and Rob is also
inertial. Suppose Alice and Rob are separated far enough, so the mutual
influences can be safely ignored again. Then, in the weak-coupling limit
with $\Omega \gg \gamma \Lambda_1 \gg a \to 0$, one has
\begin{equation}
  \Sigma \approx {\hbar^2 e^{-4\gamma t}\over 16\alpha^2\beta^2\Omega^2}
  \left[ Z_8 \left( e^{-4\gamma t} -2 e^{-2\gamma t}\right) + Z_4\right],
\label{Sig0}
\end{equation}
where
\begin{eqnarray}
  Z_8 &\equiv& \left(\hbar -\alpha^2\Omega\right)^2
    \left(\beta^2-\hbar\Omega\right)^2,\\
  Z_4 &\equiv& \hbar^2\left(\beta^4 +\alpha^4\Omega^4 +
    6\alpha^2\beta^2\Omega^2\right) - 2\hbar\Omega\left(\beta^2+
    \alpha^2\Omega^2\right)\left(\hbar^2+\alpha^2\beta^2\right) .
\end{eqnarray}
It is clear that $Z_8\ge 0$ and $Z_8-Z_4 \ge 0$. When $Z_4 >0$, the
disentanglement time is clearly finite:
\begin{equation}
  t_{dE} \approx -{1\over 2\gamma}\ln\left( 1-\sqrt{1-{Z_4\over Z_8}}\right).
\end{equation}
Indeed, one has $\gamma t_{dE} \approx 0.125$ in the right plot of
Fig. \ref{AREta0}.

When $Z_4<0$, the disentanglement time looks infinite. But again, the
corrections of the next order yields
\begin{equation}
  \Sigma \approx {\hbar^2 \over 16\alpha^2\beta^2\Omega^2}
       Z_4 e^{-4\gamma t}
    + {\hbar^3 \gamma\Lambda_1 \over 4\pi\alpha^2\beta^2\Omega^2}
    Z_2 e^{-2\gamma t} +
    {\hbar^4\over \pi^2\Omega^2} \gamma^2\Lambda_1^2
\end{equation}
around $t\approx t_{dE}\gg 1$ for large $\Lambda_1$, with
\begin{equation}
  Z_2 \equiv \alpha^2\left(\beta^2 -\hbar\Omega\right)^2 + \beta^2
    \left(\alpha^2 \Omega -\hbar\right)^2 \ge 0.
\end{equation}
This gives a finite disentanglement time
\begin{equation}
  t_{dE} \approx {1\over 2\gamma}\ln { |Z_4|\pi/(2\hbar \gamma\Lambda_1)
  \over Z_2 + \sqrt{Z_2^2-4\alpha^2\beta^2 Z_4} } ,
\end{equation}
and $\gamma t_{dE}\approx 3.96$ in Fig. \ref{AREta0} (left).

The dotted curves in Figs. \ref{AREt} (left) and \ref{AREta0} (left
and right) are those $\Sigma$s with vacuum fluctuations of the field
switched off, namely, with $\left<\right.{\cal R}_\mu, {\cal R}_\nu
\left. \right>_{\rm v}$ set to be zero. In Fig. \ref{AREt} (left) it
seems that vacuum fluctuations would reduce $|\Sigma |$ thus
suppressing the entanglement. This is not true. One can verify that,
for sufficiently large $t$, the dashed curve in Figs. \ref{AREt}
(left) will overtake the solid curve and then become positive, just
like what Fig. \ref{AREta0} (right) suggests.
In the weak-coupling limit, the vacuum fluctuations of the field that
the detectors see locally do not always suppress (or enhance) quantum
entanglement beyond the disentanglement due to the dissipation of
initial quantum fluctuations of the detectors (corresponding to the
exponential decay of $\left<\right. .. \left.\right>_{\rm a}$ in time.)

\subsubsection{Evolution of entanglement in Rob's proper time
(Rindler time)} \label{weakRin}

\begin{figure}
\includegraphics[width=18cm]{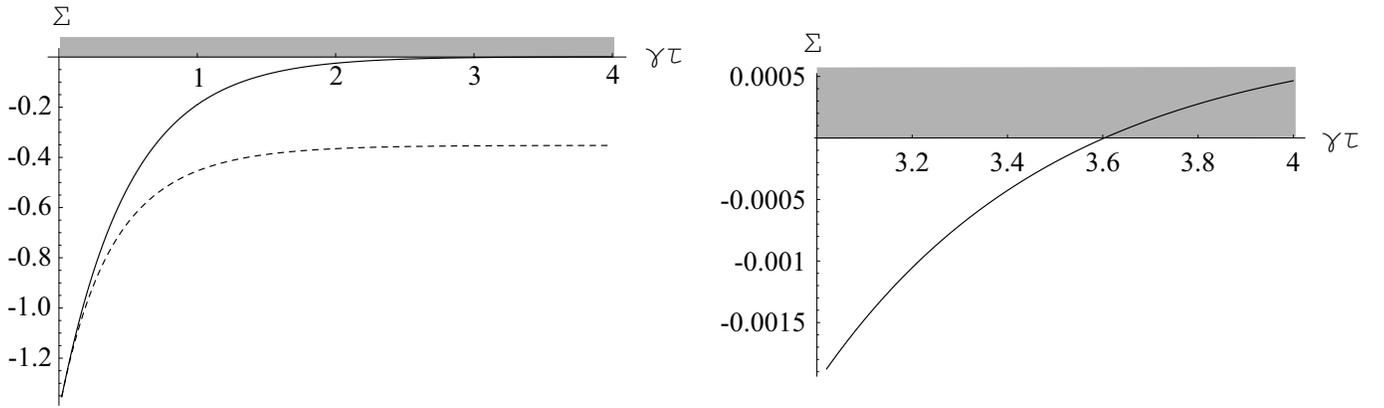}
\caption{The solid curves represent $\Sigma$ in its totality in Rob's 
point of view. ``Sudden death" of entanglement occurs at $\gamma \tau_{dE}
\approx 3.61$, as can be seen in the enlargement on the right plot.
The dotted curve represents $\Sigma$ with all $\left<\right. ..\left.
\right>_{\rm v}$ set to be zero. All parameters are the same as
those in Fig. \ref{AREt} except $\gamma=10^{-6}$, $a=2$ and $b=4.1$.}
\label{SRob}
\end{figure}

The hypersurface with constant Rindler time $\tau$ extends to $t<0$ region
in the left half of the Minkowski space (see Fig.\ref{setup}).
Suppose, even {\it before} the initial time slice with $t=0$, the
field state has 
been the Minkowski vacuum. 
Then the field state in Rindler time slicing is also 
a Gaussian state and the quantity $\Sigma(t, \tau)$, with a similar
definition to $(\ref{Entcond})$ but we let $t=b^{-1}\tanh a\tau$ here,
can still serve as an 
measure of the entanglement between Alice's and Rob's detector in Rob's frame. 

For $a$ sufficiently large, $\gamma t \to \gamma/b \ll 1$ for large
$\tau$. Then
\begin{equation}
  \Sigma(t=b^{-1}\tanh a\tau, \tau) \approx
    {\hbar^2\over 64\alpha^2\beta^2} \left(\hbar^2 -
    \alpha^2\beta^2\right)^2 \left[ \left(1-e^{-2\gamma\tau}\right)^2
    \coth^2{\pi\Omega\over a} -\left(1+e^{-2\gamma\tau}\right)^2\right],
\end{equation}
and the disentanglement time is approximately
\begin{equation}
  \tau_{dE} \approx {\pi\Omega\over \gamma a}, \label{tdERob}
\end{equation}
in Rob's point of view (see Fig. \ref{SRob}).                                
This result may be interpreted as presenting a                               
dynamical version of the statement ``quantum entanglement is degraded          %
in noninertial frames" \cite{AM03, FSM05}. Note that in Rob's frame
Alice will never cross the event horizon. So the quantum entanglement
here is not the one ``across the event horizon".

When $a$ gets smaller, $\Sigma$ approaches those described in the
previous subsections. When $a\to 0$, the value of the finite
disentanglement time recovers the disentanglement time in the case             %
of two inertial detectors.                                                   

\subsection{Beyond the ultraweak-coupling limit}

\subsubsection{High acceleration regime}

\begin{figure}
\includegraphics[width=8cm]{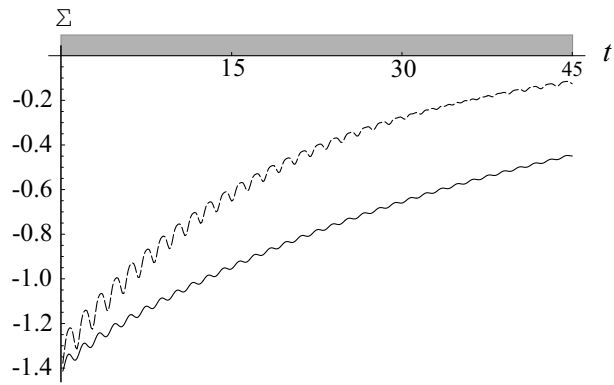}
\caption{$\Sigma$ for Rob in high acceleration (solid curve) with
$a=10$ and $b=21$. ($a=.01$, $b=.021$ for the dashed curve for
comparison.) The large time dilation makes Rob appear almost frozen
in the view of Alice at large $t$, and the disentanglement time is
longer for larger $a$. Here $\gamma = 0.01$, $\Lambda_0=\Lambda_1=50$,
and other parameters are the same as those in Fig. \ref{AREt}.  }
\label{higha}
\end{figure}

When the proper acceleration of Rob is very large, Rob will reach a
very high speed in a very short time, when the time dilation makes
Rob appear almost frozen in the view of Alice, and the disentanglement
process becomes slower than those in the case with Rob's acceleration
smaller. In other words, in our setup, the larger acceleration Rob has,
the longer the disentanglement time in Alice's clock (see Fig.
\ref{higha}), while in Rob's frame $\tau_{dE}$ is shorter from
$(\ref{tdERob})$. The statement ``a state which is maximally entangled
in an inertial frame becomes less entangled if the observers are
relatively accelerated" \cite{FSM05} is too simplistic and could be
misleading here.

\subsubsection{strong-coupling regime}                                      

When the coupling becomes larger, there will be oscillation emerging
on top of the smooth curve of $\Sigma$ in the ultraweak-coupling limit.
So around $t_{dE}$ the entanglement between the detectors of Alice and
Rob may disappear and revive for several times, before they finally become
separable forever. The duration of such kind of entanglement revival will
not exceed the order of time scale $1/\Omega$ (see the upper plot of
Fig. \ref{dEvsd}.)

\begin{figure}
\includegraphics[width=8cm]{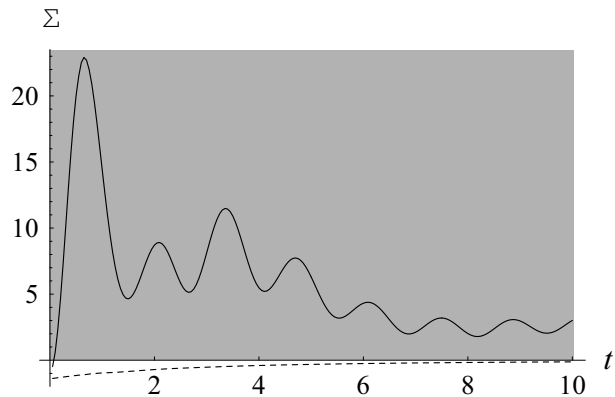}
\caption{$\Sigma$ in the strong-coupling regime. The entanglement between
Alice and Rob in their initial state is annulled by the strong impact of
vacuum fluctuations ($\Sigma >0$) right after the coupling is switched on.
Here $\gamma = 0.1$, $\Lambda_0=\Lambda_1=50$, $a=1$, $b=2.01$, and other
parameters are the same as those in Fig. \ref{AREt}. The dotted
curve at the bottom are the $\Sigma$ with all $\left<\right. .. \left.
\right>_{\rm v}$ turned off. Numerically we find that the higher order
corrections to $\Sigma$ from mutual influences are less than $2 \% \sim
 O(\gamma^2)$.}
\label{strongcoup}
\end{figure}

When the coupling gets even larger, the vacuum fluctuations will             
exert a strong impact on quantum entanglement.                               
Usually this shortens the disentanglement time.
Indeed, in Fig. \ref{strongcoup} we find that, for $\gamma\Lambda_1$
sufficiently large, the initial quantum entanglement is annulled
right after the coupling is switched on.

The higher order corrections to $\Sigma$ from retarded mutual
influences of two detectors, while their amplitudes decay in time,
are not always positive or negative. Nevertheless, the memory effect
induced by mutual influences contributes little 
in the cases considered: Although the total corrections to
$\Sigma$ from mutual influences become more obvious for a larger
coupling, they are of $O(\gamma^2)$ and remain small as seen in Fig.
\ref{strongcoup}.
Results with stronger mutual influences will be reported in a future work.   

\section{Discussion}

\subsection{Infinite disentanglement time in Markovian limit}

A criterion in the Markovian limit on finite disentanglement time has
been offered by Yu and Eberly \cite{YE04}. In some parameter range
($a \in (0, 1/3)$ there) the concurrence decays exponentially in time
so the disentanglement time looks infinite. Here we have similar
situations in the ultraweak-coupling limit as discussed in Secs.
\ref{weakMin} and \ref{2inertMin}. In particular, the parameter $Z_4$
in $(\ref{Sig0})$ plays an analogous role to $a$ in \cite{YE04}.
While the entanglement gets a clear sudden death when $Z_4 > 0$, the
disentanglement time looks infinite otherwise. However, in Secs.
\ref{weakMin} and \ref{2inertMin} we showed that the correction from
the next order to the ultraweak-coupling approximation will render
the disentanglement time of the latter case finite. According to our
results, it will be interesting to see whether the infinite
disentanglement time in the Markovian cases that Yu and Eberly
considered would become finite if one considers the corrections beyond the
Markovian approximation.

\subsection{Entanglement and correlation}

When cross correlations vanish, two detectors are uncorrelated and
the two-point correlation matrix ${\bf V}$ defined in
$(\ref{CorrMtx})$ is block diagonalized so the Wigner function, or
the reduced density matrix, for the detectors can be factorized into
a tensor product of two Wigner functions or reduced density matrices for
each detector. Now two detectors are {\it simply separable}, with a
stronger condition than those for disentangled or {\it separable}
states.

In our model, the cross correlations $\left<\right.{\cal R}_A, {\cal
R}_B \left.\right>$ always vanish as $\gamma t \to \infty$, while the
two-point functions of each detector remain finite. We found
$|\left<\right.{\cal R}_A, {\cal R}_B\left.\right>| \sim e^{-\epsilon
t}$ with $\epsilon \propto \gamma$ at sufficiently large $t$.
This implies that there is no residual entanglement at late times
because
\begin{equation}
  \Sigma|_{\gamma t\to \infty} =  \left.\left[ \left<\right.Q_A^2
  \left.\right>^{(0)}\left<\right.P_A^2\left.\right>^{(0)} -
  {\hbar^2\over 4}\right] \left[ \left<\right.Q_B^2\left.\right>^{(0)}
  \left<\right.P_B^2\left.\right>^{(0)}- {\hbar^2\over 4}
  \right]\right|_{\gamma t\to \infty},
\end{equation}
which is a product of the uncertainty relations for each detector in
steady state and is positive-definite if the coupling $\lambda_0$ is
nonvanishing. (Note that $\left<\right.Q_A, P_A\left.\right> \sim
e^{-\gamma t} $ and $\left<\right.Q_B, P_B\left.\right> \sim \exp
[-(a+\gamma)a^{-1} \sinh^{-1} a t]$ vanish at late times; for
explicit expressions of other late-time two-point functions, see
Appendix A in Ref. \cite{LH2006}.) In Minkowski time, two detectors
become separable after a finite disentanglement time $t_{dE}$, but
not simply separable or uncorrelated until $\gamma t \to \infty$.

\subsection{time slicing and coordinate dependence}                       
Consider two events, one in Alice's world line at her                       %
proper time $t=\bar{t}$, the other in Rob's world line at his proper
time $\tau=\bar{\tau}$, defined on the same time slice in some
coordinate. Recall that the physical RDM for the two detectors are
obtained by integrating out the degrees of freedom of the field
defined on the same hypersurface associated with a certain time          
slicing. Thus the RDM of the two detectors at the two events
$z^\mu_A(\bar{t})$ and $z^\mu_B(\bar{\tau})$ should always be
associated with a specification of a time slicing scheme, so should       
the criterion of separability derived from the RDM. But there are
infinitely many choices of time slices intersecting Alice and Rob's
world line right at these two events. Does quantum entanglement of
the detectors at $z^\mu_A(\bar{t})$ and $z^\mu_B(\bar{\tau})$ depend
on different choices of time slice?

In the cases we considered in this paper, the answer is no. In our UD
detector theory, the coordinate transformation from Minkowski
coordinate to a new coordinate consistent with some other
time slicing scheme such as Rindler time will not change the
quadratic form of the action $(\ref{Stot1})$, so the combined system
in the new coordinate is still linear. If the initial time slice in
the new time slicing scheme is identical to our $t=0$ hypersurface
in Minkowski coordinate, and the combined system is starting with a
Gaussian initial state such as $(\ref{initstat})$, the quantum state
of the combined system will remain Gaussian during the evolution in     
the new time slicing scheme. Hence the RDM of the detectors is still    
Gaussian and the criterion $(\ref{Entcond})$ can be applied in this     
new time slicing. The Gaussianity implies that both the RDM of the
detectors and the criterion $(\ref{Entcond})$ are fully described by
the two-point functions of the detectors $\left<\right.{\cal R}_\mu ,   
{\cal R}_\nu \left.\right>$, which are independent of the choice of
time slice connecting these two events. Thus the RDM and the
separability of the Gaussian state at those two events are
independent of time slicing.  We believe that this is a general
feature of quantum entanglement in relativistic quantum information
theory.

Nevertheless, in Secs.\ref{weakMin} and \ref{weakRin} we learnt that the
entanglement {\it dynamics} and the disentanglement (proper) time are
coordinate-dependent, because in general two events simultaneous in one
({\it e.g.} Minkowski) coordinate are not simultaneous in another
({\it e.g.} Rindler) coordinate, while quantum entanglement is a property
of a quantum state of two events at the same time slice.
\footnote{One may wonder how the criterion $(\ref{Entcond})$ fares with        
time-like separated events such as obtained by inserting some $(t, \tau)$        %
to $\Sigma$, {\it e.g.} Alice at $t=10$ and Rob at $\tau=10^{-6}$ with $a=1$     %
and $b=2.1$. But this is too much of a stretch because timelike separated        %
events will not be on the same time slice.}                                    



\subsection{Detector-detector entanglement vs detector-field
entanglement}

\begin{figure}
\includegraphics[width=12cm]{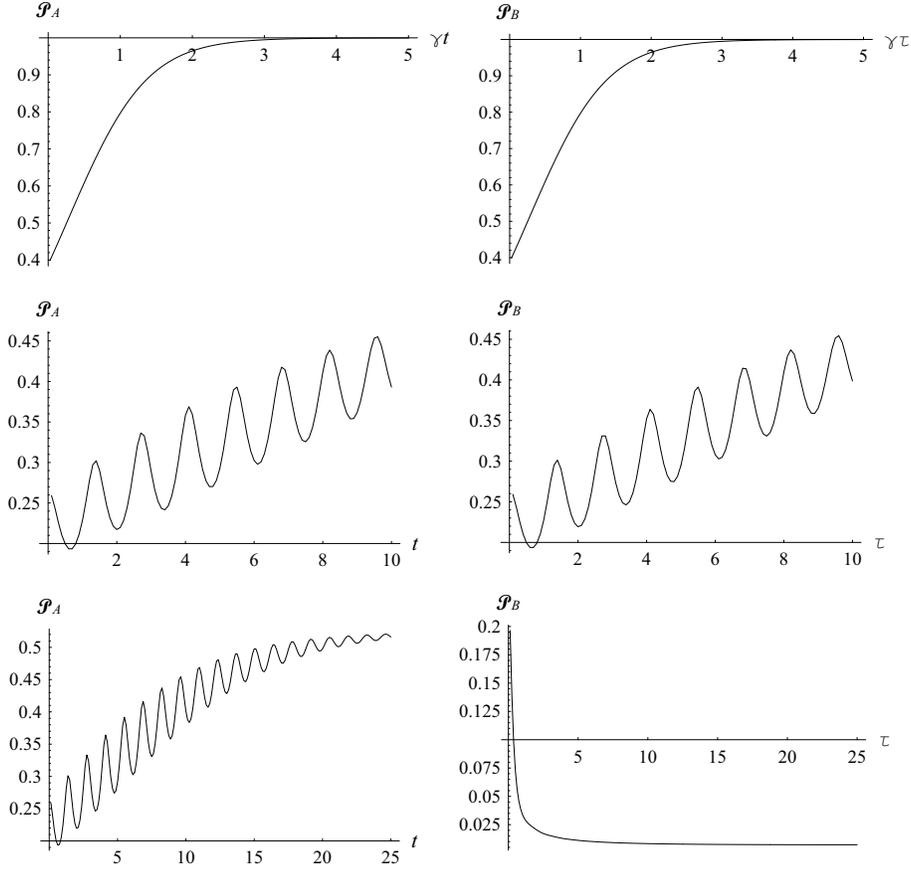}
\caption{Purities of Alice (${\cal P}_A$) and Rob (${\cal P}_B$).
(Top) ${\cal P}_A$ and ${\cal P}_B$ in the ultraweak-coupling regime,
with the same parameters as those in Fig. \ref{AREt}. In this regime
${\cal P}_A$ and ${\cal P}_B$ each in its own proper time has
virtually the same behavior.
(Middle) ${\cal P}_A$ and ${\cal P}_B$ in the strong-coupling regime, with        
the same parameters as those in Fig. \ref{strongcoup}. Again their
difference is very tiny.
(Bottom) ${\cal P}_A$ and ${\cal P}_B$ in high acceleration limit, with
the same parameters as those for the solid curve in Fig. \ref{higha}.
Now their difference is evident.}
\label{cohe}
\end{figure}

Quantum coherence in a detector is related to its purity, which also
indicates the degree of quantum entanglement between that detector and the
rest of the combined system (the field and the other detector)\cite{BZ06}.
Since the combined system is a Gaussian state, the reduced density matrix
for each detector is also Gaussian. So the purities of Alice and
Rob's detectors are simply \cite{LH2006}
\begin{equation}
  {\cal P}_j = {\hbar /2 \over \sqrt{\left<\right.P_j^2\left.\right>
    \left<\right.Q_j^2\left.\right> - \left<\right.P_j, Q_j\left.\right>^2}},
\end{equation}
where $j=A,B$. Obviously the information contained in the cross
correlations $\left<\right.{\cal R}_A, {\cal R}_B\left.\right>$ is
ignored in every combination of ${\cal P}_j$. We illustrate some
examples in Fig. \ref{cohe}. One can see that the evolution of purity
in each detector is quite different from the evolution of
entanglement between them. There is no ``sudden death" of quantum
entanglement between one detector and the rest, and the late-time
purity is always less than one. This is a consequence of the direct
interaction between each detector and the field.

\subsection{Disentanglement time and separation of detectors}

\begin{figure}
\includegraphics[width=12cm]{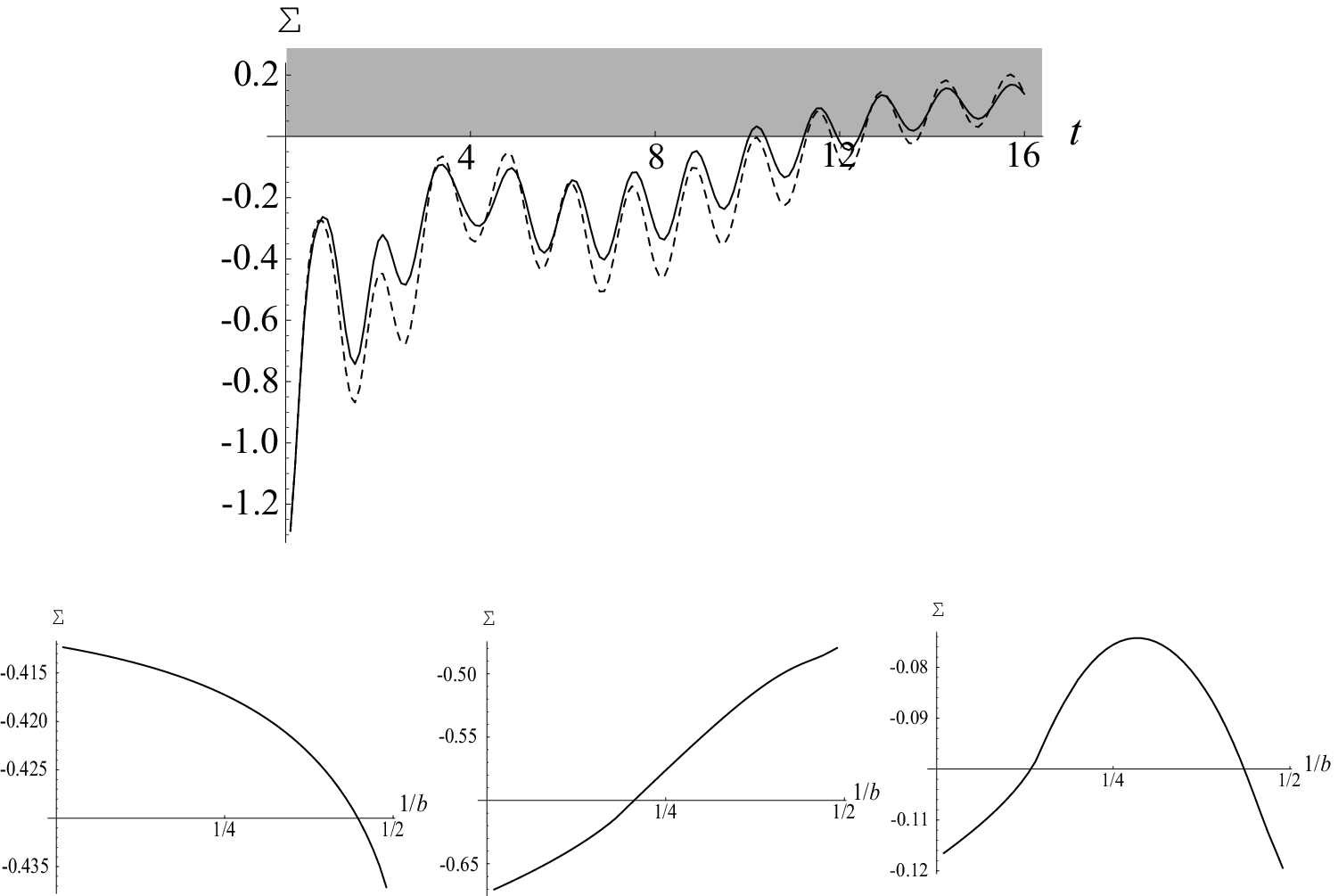}
\caption{(Upper) $\Sigma$ with initial separation $d_I\approx 0.5025$
($b=2.01$) for the solid curve and $d_I=0.9999$ ($b=10000$) for the
dashed curve. Alice is located at $z_A^1 = 1/b$ and the initial
separation between Alice and Rob at $t=0$ is $d_I = a^{-1}-b^{-1}$.
Here $a=1$, $\gamma =0.025$, $\Lambda_0=\Lambda_1=50$, and other parameters
are the same as those in Fig. \ref{AREt}. One can see that short-time
revivals of entanglement occur during $t\approx 10\sim 11$ and around $t=12$.
(Lower) The dependence of $\Sigma$ on the initial separation is not evident
systematically: from left to right are $\Sigma$ at $t=0.5, 2.5, 3.2$,
respectively. }\label{dEvsd}
\end{figure}

Naively one may think that spatially the closer Alice and Bob are, the
larger $\left<\right. {\cal R}_A, {\cal R}_B \left.\right>$, so the
disentanglement time gets shorter. Our results in Fig. \ref{dEvsd}
show that this is not true. One can hardly see any simple relation
between the initial separation of the two detectors and the
disentanglement time. The larger the separation, the stronger the
entanglement ($\Sigma$ more negative) at some moments,
but weaker at others.

We have considered the case with both detectors being at rest and
spatially separated. While the higher order corrections from mutual
influences are complicated and quite hard to handle as $t$ increases,
for separation $d$ large enough that the mutual influences cannot
reach in time (namely, $d > c t$), we get a similar result that no
simple proportionality exists between the separation and the degree
of entanglement of the two detectors. As the system evolves,              
sometimes the entanglement is stronger for larger separation,
sometimes it is weaker.

\subsection{How generic are the features contained in this model?}

Our results show that the disentanglement time in our model is
finite in all cases considered. But how generic are these results?

Our model has two UD detectors spatially separated in (3+1)D spacetime,
one stationary and another uniformly accelerating and running away,
without any direct interaction with each other.
The presence of the event horizon for Rob has the effect of curtailing         
the mutual influences higher than a certain order.                             
When we focus on one single UD detector, it behaves like the
QBM of a harmonic oscillator interacting
with an Ohmic bath provided by the (3+1)D scalar quantum field
\cite{HM94}.

If there exists a direct interaction between the two detectors, there
will certainly be residual entanglement between them at late times,
just like the late-time entanglement between each detector and the
field in our model.

The scalar field in our model is living in a free space. The presence
of boundaries will change the result. The backreaction started from
each detector when reflected by boundaries will add another memory
effect to the dynamics of entanglement and affect the late-time
behavior of detectors. One example is an electromagnetic field in a
perfect cavity. Spacetime dimension also matters. Different spacetime
dimensions or boundaries give different spectral density functions of
the field experienced by the detectors \cite{HM94}. Moreover, in
$(n+1)$D, $n\ge 3$ the spatial separation $d$ implies a suppression of
mutual influences by a factor $1/d^{n-2}$, so the mutual influences
would be negligible in the cases with large spatial separation.

Spatial separation also contributes to the retardation of mutual
influences, and since Alice is at rest and Rob is in uniform acceleration,
one detector's dynamics is generally out of phase from the backreaction
of the other. If the two detectors are located very close to
each other through the whole dynamical process, or the two detectors are
separated at a distance where resonance ($d\approx 2\pi c /\Omega$) is
set up, the features of the disentanglement process can be different.
For example, residual entanglement at late times under specific
conditions in the 2HO QBM model has been reported in \cite{PR07, AZ07};
that model is equivalent to a detector theory with two UD detectors
sitting at the same point in 3D space, when the HO bath is ohmic.

Therefore, we believe that our results in this paper are
generic for two well-separated detectors or atoms without direct
interaction with each other, but coupled to a common quantum field in
$(n+1)$D, $(n\ge 3)$ Minkowski space without boundary.\\

\noindent{\bf Acknowledgement} SYL wishes to thank Jun-Hong An for
illuminating discussions, and David Ostapchuk for finding a small localized         
mistake in the calculation reported in the first version which does not               
affect any claims or results reported.                                                
BLH and SYL thank Rafael Sorkin for a useful comment on the dependence of             %
the Gaussianity of a state on slicing.                                              
This work is supported in part by NSF Grants No. PHY-0601550 and No. PHY-0426696.

\begin{appendix}

\section{Expressions for mode functions}

For the cases with $b>2a$, one has
\begin{eqnarray}
  & & q_A^{(A)}(t) = \theta(t) {e^{-\gamma t}\over 2}
    \left[W_- e^{i\Omega t}+ W_+ e^{-i\Omega t}\right]
    \nonumber\\ & & \,\,\, + {\theta \left(t - t_2\right)
    \lambda_0^4\over 16\pi^2 a \Omega^2} \int^t_{t_2}
    {dt' K(t-t') \over t'^2-b^{-2}+a^{-2}}
    \int^{\tau_-(t')}_{\tau_1} {d\tau' K[\tau_-(t')-\tau']
    \over a^{-1}\cosh a\tau' -b^{-1}} 
    e^{-\gamma t_-(\tau')} \left[ W_- e^{i\Omega t_-(\tau')}
    +W_+ e^{-i\Omega t_-(\tau')} \right], \label{qAA}\\
  & & q_B^{(A)}(\tau) = \theta\left(\tau-\tau_1\right)
    {\lambda_0^2\over 8\pi\Omega}\int^\tau_{\tau_1}
    {d\tau' K(\tau-\tau') \over a^{-1}\cosh a\tau' - b^{-1}}
    e^{-\gamma t_-(\tau')} \left[W_- e^{i\Omega t_-(\tau')}
    +W_+ e^{-i\Omega t_-(\tau')}\right],\\
  & & q_A^{(B)}(t) = \theta\left(t - t_1\right)
    {\lambda_0^2\over 4\pi a \Omega} \int^t_{t_1} {dt' K(t-t')
    \over t'^2-b^{-2}+a^{-2}} e^{-\gamma\tau_-(t')}
    \left[W_-e^{i\Omega\tau_-(t')}+ W_+ e^{-i\Omega\tau_-(t')}\right],\\
  & & q_B^{(B)}(\tau) = \theta(\tau) {e^{-\gamma\tau}\over 2}\left[
    W_- e^{i\Omega\tau}+ W_+ e^{-i\Omega\tau}\right],\\
  & & q_A^{(+)}(t,{\bf k}) = \theta(t){\lambda_0\over\Omega}\int_0^t dt'
    K(t-t') f_0^{(+)}(z_A(t'),{\bf k})\nonumber\\
    & & \,\,\, +\theta\left(t-t_1\right){\lambda_0^3\over 2\pi a \Omega^2}
    \int^t_{t_1}{dt' K(t-t')\over t'^2- b^{-2}+ a^{-2}}
    \int_0^{\tau_-(t')}d \tau' K[\tau_-(t')-\tau']
    f_0^{(+)}(z_B(\tau'),{\bf k}) \nonumber\\
    & &\,\,\,+{\theta\left(t-t_2\right)\lambda_0^5\over 8\pi^2 a\Omega^3}
    \int^t_{t_2}{dt'K(t-t') \over t'^2-b^{-2}+a^{-2}}
    \int^{\tau_-(t')}_{\tau_1} {d\tau' K[\tau_-(t')-\tau']
    \over a^{-1}\cosh a\tau'-b^{-1}}
    \int_0^{t_-(\tau')}dt''
    K[t_-(\tau')-t'']f_0^{(+)}(z_A(t''),{\bf k}),\\
  & & q_B^{(+)}(\tau, {\bf k}) = \theta(\tau){\lambda_0\over\Omega}
    \int_0^\tau d\tau' K(\tau-\tau')
    f_0^{(+)}(z_B(\tau'),{\bf k})\nonumber\\
    & &\,\,\,+\theta\left(\tau-\tau_1\right){\lambda_0^3\over 4\pi\Omega^2}
    \int^\tau_{\tau_1}{d\tau' K(\tau-\tau')\over a^{-1}\cosh a\tau'-b^{-1}}
    \int_0^{t_-(\tau')}
    dt' K[t_-(\tau')-t']f_0^{(+)}(z_A(t'),{\bf k}). \label{qBp}
\end{eqnarray}
where $\tau_1 \equiv a^{-1}\ln(b/a)$, $t_1=a^{-1}-b^{-1}$,
$t_2 \equiv ba^{-2}-b^{-1}$ (see Fig. \ref{setup}),
$\tau_-(t')=a^{-1}\ln a\left(t' + b^{-1}\right)$,                             
$t_-(\tau')\equiv b^{-1}-a^{-1}e^{-a \tau'}$,
$W_\pm \equiv 1\pm [(\Omega_r+i\gamma)/\Omega]$, and
$K(x)\equiv e^{-\gamma x}\sin\Omega x$.
In ultraweak-coupling limit, we shall neglect $O(\lambda_0^2)$ terms.

\section{Two detectors at rest weakly coupled to a thermal bath}

In the ultraweak-coupling limit, for $Q_A$ and $Q_B$ both inertial and in
contact with a thermal bath of temperature $T$, one has
\begin{eqnarray}
  & & \left<\right. Q_j^2(\tau_j)\left.\right>_{\rm v} \approx
    {\hbar\over 2\Omega}\coth{ \Omega\over 2T}\left( 1-e^{-2\gamma\tau_j}
    \right), \,\,\,\,\, \left<\right. P_j^2(\tau_j)\left.\right>_{\rm v}
    \approx \Omega^2 \left<\right. Q_j^2(\tau_j)\left.\right>_{\rm v} ,
    \label{Qj2vweak}\\
  & & \left<\right. Q_j^2(\tau_j)\left.\right>_{\rm a} \approx
    e^{-2\gamma \tau_j} \left( c_1^+ \cos^2\Omega \tau_j +
    c_2^+ \sin^2\Omega \tau_j \right), \label{Qj2aweak} \\
  & & \left<\right. P_j^2(\tau_j)\left.\right>_{\rm a} \approx \Omega^2
    e^{-2\gamma \tau_j}\left( c_1^+ \sin^2\Omega \tau_j +
    c_2^+ \cos^2\Omega \tau_j \right), \\
  & & \left<\right. P_j(\tau_j),Q_j(\tau_j)\left.\right>_{\rm a} \approx
    \Omega e^{-2\gamma \tau_j}\left( c_2^+ -c_1^+\right)
    \sin\Omega \tau_j\cos\Omega \tau_j,\\
  & & \left<\right. Q_A(t),Q_B(\tau)\left.\right>_{\rm a} \approx
    e^{-\gamma (t+\tau)} \left( c_1^- \cos\Omega t \cos\Omega\tau +
    c_2^- \sin\Omega t\sin\Omega\tau \right),\\
  & & \left<\right. P_A(t),P_B(\tau)\left.\right>_{\rm a} \approx \Omega^2
    e^{-\gamma (t+\tau)} \left( c_1^- \sin\Omega t \sin\Omega\tau +
    c_2^- \cos\Omega t\cos\Omega\tau \right),\\
  & & \left<\right. P_A(t),Q_B(\tau)\left.\right>_{\rm a} \approx \Omega
    e^{-\gamma (t+\tau)} \left( c_2^- \cos\Omega t \sin\Omega\tau -
    c_1^- \sin\Omega t\cos\Omega\tau \right),\\
  & & \left<\right. Q_A(t),P_B(\tau)\left.\right>_{\rm a} \approx \Omega
    e^{-\gamma (t+\tau)} \left( c_2^- \cos\Omega\tau\sin\Omega t -
    c_1^- \sin\Omega \tau \cos\Omega t \right), \label{QAPBaweak}
\end{eqnarray}
where $j=A,B$, and
\begin{eqnarray}
  c_1^{\pm}&\equiv&{1\over 4}
      \left( {\hbar^2\over\beta^2}\pm\alpha^2\right),\\
  c_2^{\pm}&\equiv&{1\over 4\Omega^2}
       \left( \beta^2\pm {\hbar^2\over\alpha^2}\right).
\end{eqnarray}
Then the quantity $\Sigma$ reads
\begin{equation}
  \Sigma \approx {1\over 16 \Omega^2}
    \left(X_4 e^{-4\gamma t}+X_2 e^{-2\gamma t} +X_0\right)
    \left(Y_4 e^{-4\gamma t}+Y_2 e^{-2\gamma t} +Y_0\right),
\end{equation}
where
\begin{eqnarray}
  X_4 &\equiv& \left(\hbar -\alpha^2\Omega \coth {\Omega\over 2T}\right)
    \left(\hbar\Omega -\beta^2 \coth {\Omega\over 2T}\right),\\
  X_2 &\equiv& \left(\hbar\beta^2 + \hbar\alpha^2\Omega^2- 2\alpha^2\beta^2
    \Omega \coth {\Omega\over 2T} \right) \coth {\Omega\over 2T},\\
  X_0 &\equiv& \alpha^2\beta^2\Omega\left(\coth^2 {\Omega\over 2T}-1\right),
\end{eqnarray}
and $Y_n=X_n|_{\alpha\to (\hbar/\beta), \beta\to (\hbar/\alpha)}$, $n=0,2,4$.
For $T>0$, if $\alpha^2\beta^2\not= \hbar^2$, $\Sigma < 0$ at $t=0$ and
$\Sigma > 0$ as $t\to \infty$, so the disentanglement time $t_{dE}$ must be
finite. It is easy to verify that there exists only one solution for
$\Sigma(t_{dE})=0$: For $\alpha^2\beta^2 > \hbar^2$,
\begin{equation}
  t_{dE} \approx -{1\over 2\gamma}\ln {-X_2-\sqrt{X_2^2-4X_0 X_4}\over 2X_4},
\end{equation}
and the disentanglement time for $\alpha^2\beta^2 < \hbar^2$ is
$t_{dE}(X_n \to Y_n)|_{\alpha^2\beta^2 > \hbar^2}$.
At high temperature limit $T\to \infty$, we have $t_{dE}\sim T^{-1}$.

\end{appendix}

\end{document}